\documentclass[%
 reprint,
 amsmath,amssymb,
 aps,
]{revtex4-1}
\usepackage{graphicx}
\usepackage{dcolumn}
\usepackage{bm}
\usepackage[usenames]{color}
\usepackage{float}
\usepackage{pstricks}
\usepackage[english]{babel}
\usepackage[latin1]{inputenc}
\usepackage[usenames]{color}
\usepackage{hyperref}
\usepackage{subfigure}
\usepackage{caption}
\usepackage{mathrsfs}
\usepackage{multirow}
\usepackage{setspace}

\begin{document}
\preprint{HEP-BUAP-01.13}

\title{Analytical Calculation of Radiative Corrections of a THDM Potential}

\author{Enrique D\'iaz M\'endez}
\affiliation{%
Facultad de Ciencias F\'isico-Matem\'aticas\\
Benem\'erita Universidad Aut\'onoma de Puebla, C.P. 72570, Puebla, Pue., M\'exico.}%

\date{\today}

\begin{abstract}
We obtain a closed form effective potential at the one-loop level of a Two Higgs Doublet Model. Through the loop expansion we reproduce the expression presented by Weinberg and Coleman, showing explicitly every step involved in the calculation. The formalism is then extended to include interaction terms between two scalar doublets and a final expression of the one-loop level contributions is presented. 
\end{abstract}

\maketitle

\section{\label{sec:level1}Introduction}

In the Standard Model (SM) spontaneous symmetry breaking is performed making use of tree level terms in the potential, but this is only an approximate method in the full quantum theory, since it disregards the quantum corrections coming from virtual processes at the loop level, which are particularly important when considering zero and finite temperature studies.  In Ref. \cite{Coleman:1973jx} a first effort to extend the potential considered taking into account the one-loop level corrections, in order to study the phase transition between SSB and non-SSB scenarios for different models: the authors used explicitly the Landau gauge for the scalar electrodynamics theory. Later on, Jackiw \cite{Jackiw:1974cv} made the calculation in an arbitrary gauge up to two-loop level for a set of scalars with an \emph{O(n)} internal symmetry. Subsequently, Arnold et al. \cite{Arnold} extended the result for the whole SM with a single Higgs doublet including the thermal contributions.

However, as far as we know, no one has determined from first principles the one-loop level vacuum contributions in which more than one scalar field are present and have interaction terms, such as in the Two Higgs Doublet Model (THDM). Although different models have been presented in which several non interacting scalar fields are present. In the cosmological arena the scalar sector effective potential is useful to study the electroweak phase transition \cite{Caldwell:2013mox}, on which both vacuum and finite-temperature contributions must be taken into account.

In a recent paper \cite{Chakraborty:2015raa} a first attempt was made to have an effective potential in which two scalar fields have coupling terms present in the potential. Unfortunately the way in which multiple scalar potentials have been constructed makes use of a generic expression coming from the single scalar  one-loop contribution, without an explicit and detailed derivation.
In the present work, we obtain an effective potential for the scalar sector of a generic THDM. We first perform a detailed analysis for the single massless and massive field theories and then work out the THDM extension, we do this only for the scalar sector, since all other sectors are equal to those of the Minimal Standard Model, for which the effective potential has been obtained even up to the three loop level in some scenarios \cite{Martin:2013gka}. \newline

The rest of our paper is structured as follows: In sec. \ref{sec:derivation} we present the general features used in the derivation of the effective potential. And then proceed to calculate the massless self interacting real scalar field potential in sec. \ref{sec:Massless}. We then work out the case of a massive self interacting real scalar field potential in sec. \ref{sec:massive}, and in sec. \ref{THDM} we obtain the expression for the THDM effective potential. Our conclusions are presented in sec. \ref{conclusion}.

\section{\label{sec:derivation}Derivation of the Effective Potential}

\begin{figure}
\centering
    \subfigure[ ]{\includegraphics[scale = 0.10,angle=0] {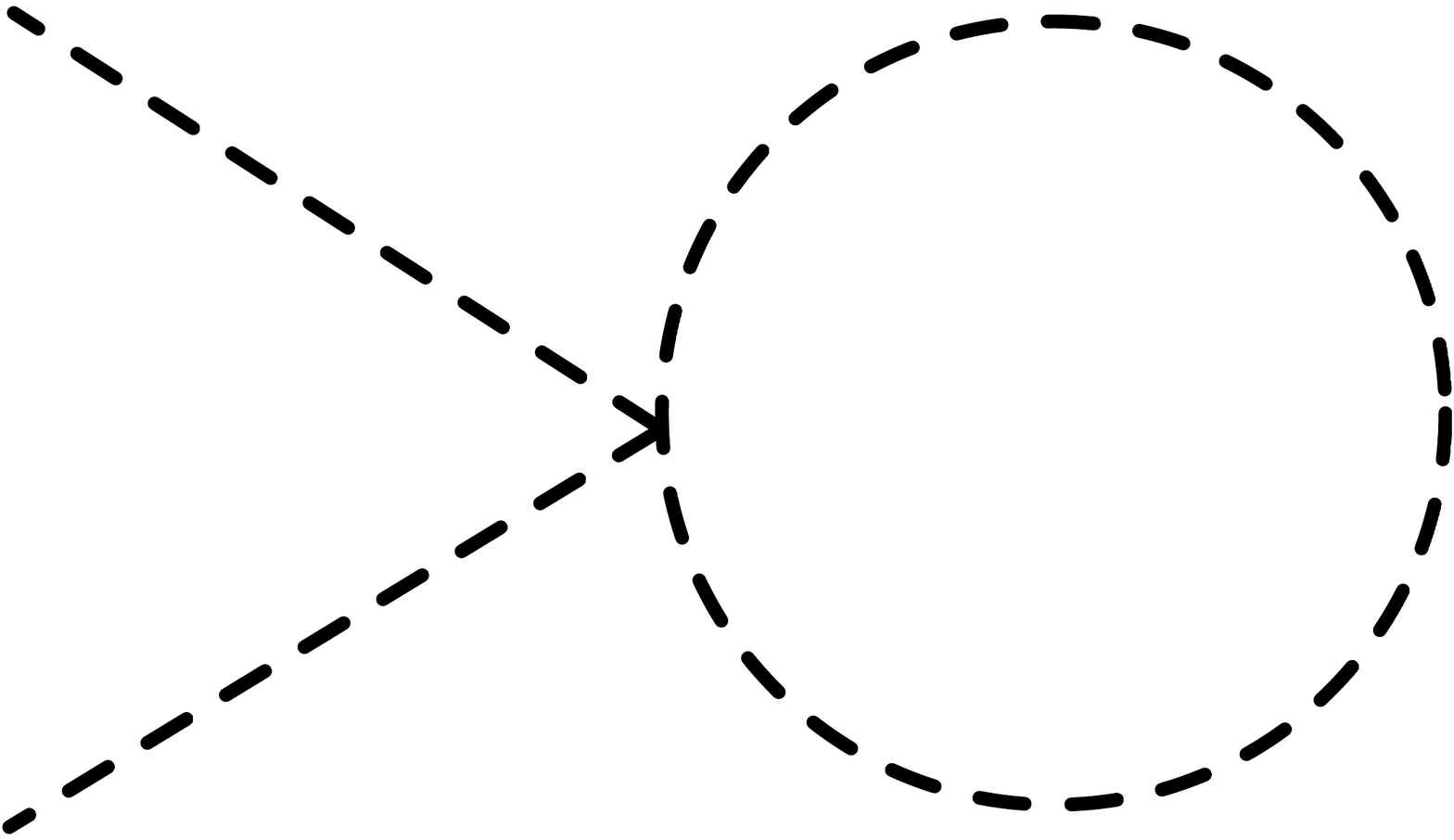}}\label{fig:a}
    \subfigure[ ]{\includegraphics[scale = 0.11,angle=90]{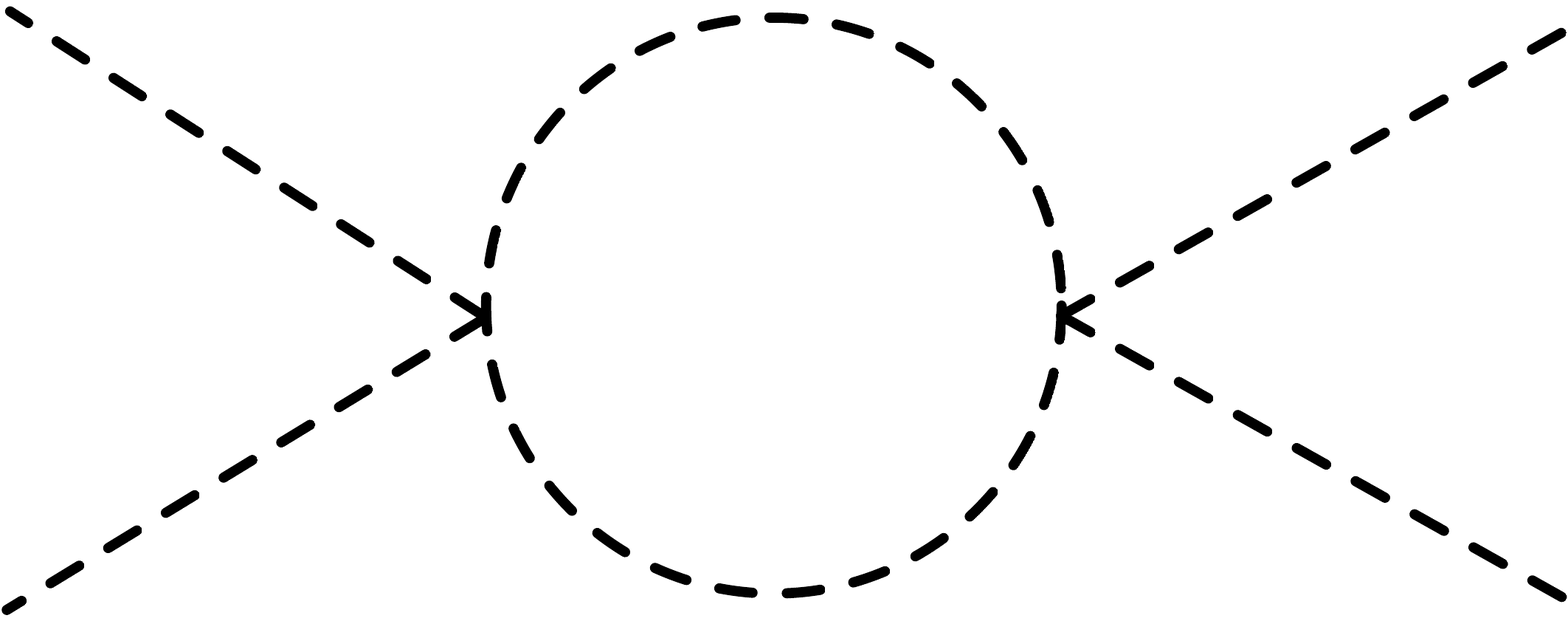}}\label{fig:b} 
    \subfigure[ ]{\includegraphics[scale = 0.10,angle=0]{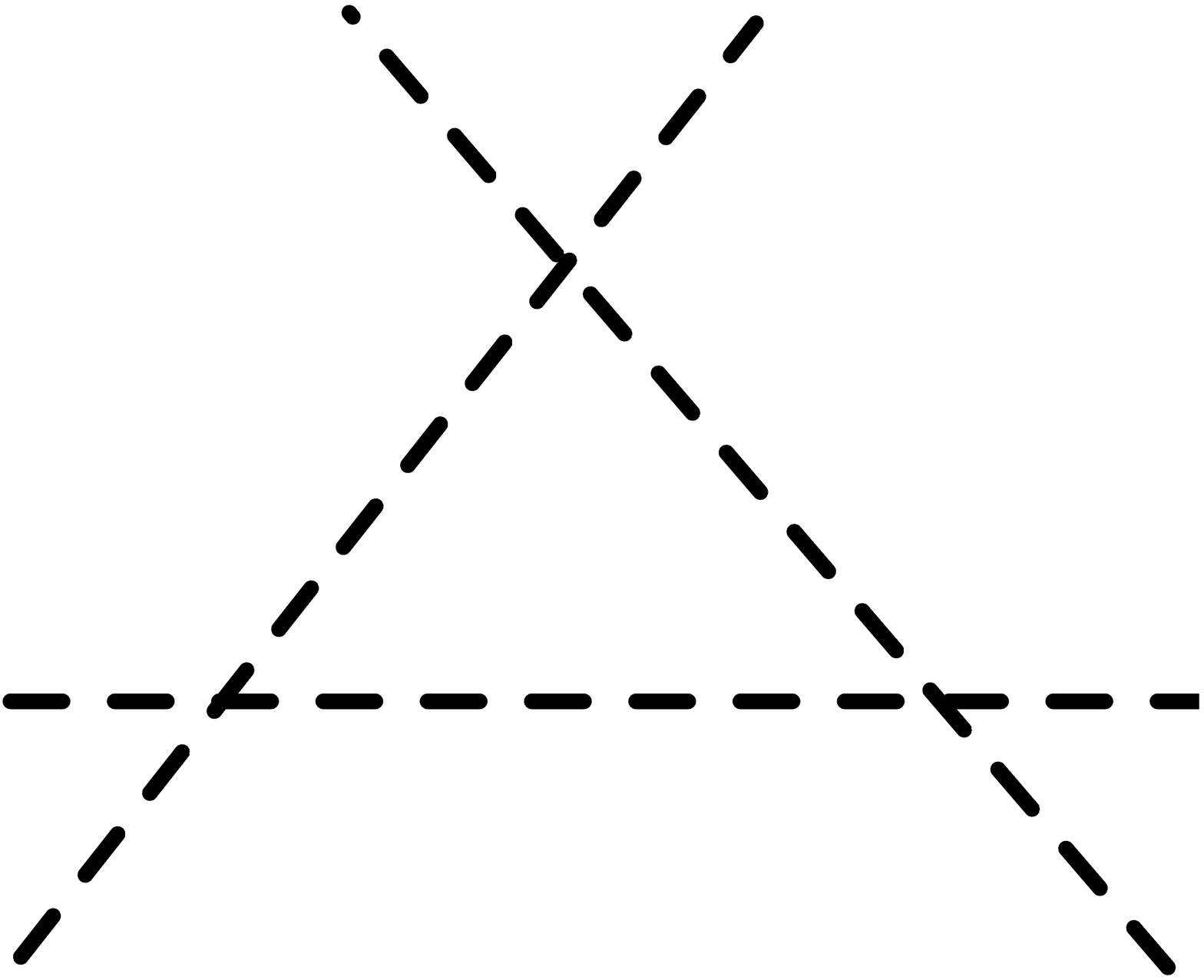}}\label{fig:c}\\
    \subfigure[ ]{\includegraphics[scale = 0.10,angle=0]{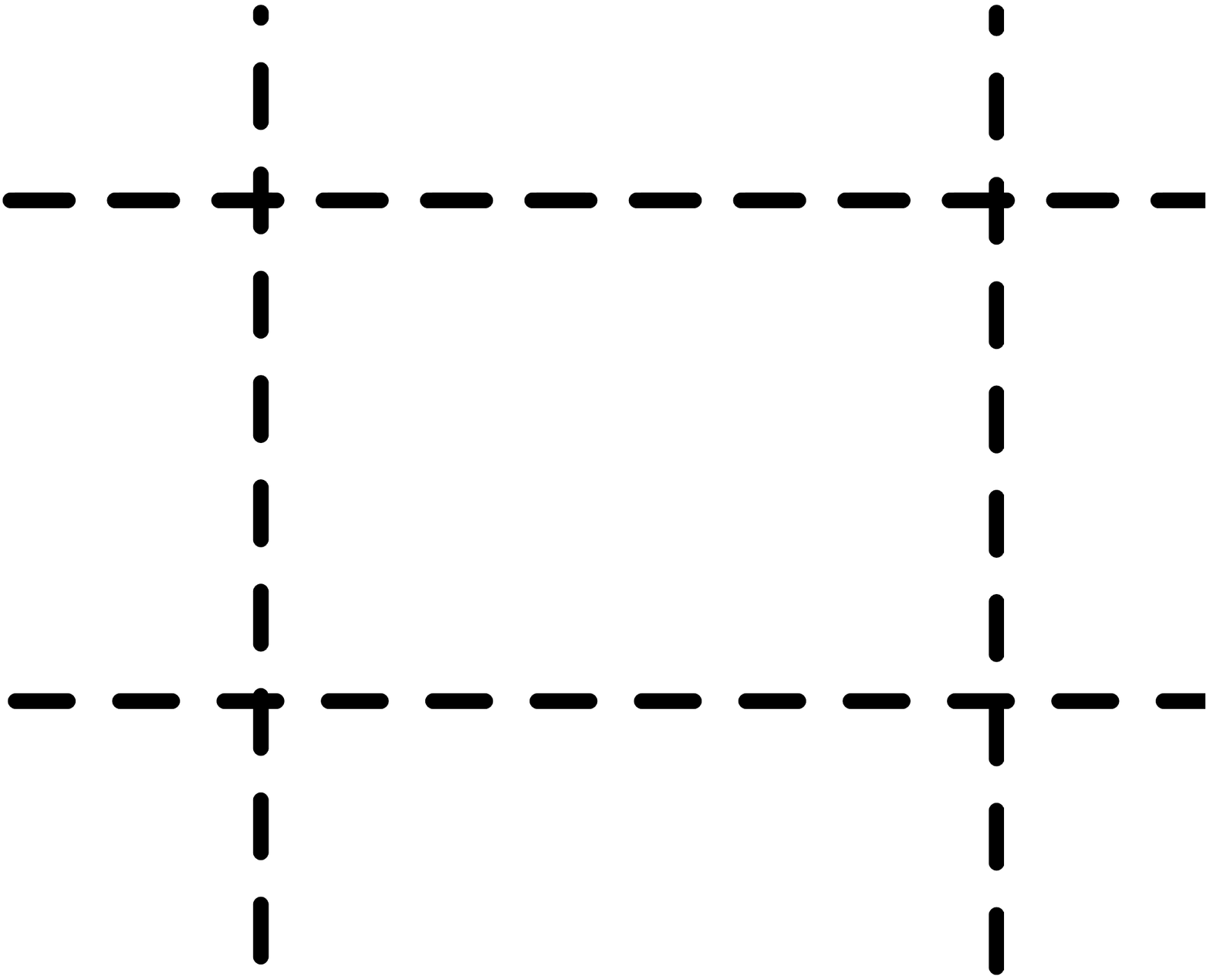}}\label{fig:d}
   \caption{First four one-loop Feynman diagrams that contribute to the effective potential. \label{fig:loop-a-d}}
	\end{figure}

For a single real massless self interactive scalar, whose Lagrangian is given by

\begin{equation}
\mathcal{L} = \frac{1}{2} (\partial_\mu \phi)^2 - \frac{\lambda}{4!}\phi^4,
\end{equation}
the one loop level corrections to the potential are given by an infinite series of Feynman diagrams; the first four of which are shown in fig. \ref{fig:loop-a-d}

 The first diagram contains only one vertex, so we only need one factor of $\lambda$ (the factor of $i$ cancels with the one of the propagator); and it has a single propagator inside the loop. The second diagram, has two vertices and two propagators. In general, the $n$-th diagram contains $n$ vertices and $n$ propagators, each one corresponding to a side of the polygon formed in the center of each diagram. 
 
 So, from the first diagram we obtain:
\begin{equation}
	\lambda\frac{1}{k^2+i\epsilon} ,
\end{equation}
where $k$ represents the momentum of the particle going around the loop. It is a virtual particle, so we actually have to integrate over all of its possible values, that is, from minus infinity to plus infinity (for each component):
\begin{equation}
	\int\frac{d^4k}{(2\pi)^4}\frac{\lambda}{k^2+i\epsilon} .
\end{equation}
The denominator $(2\pi)^4$ is conventional, and the integration limits are usually left implicit when they are $\pm\infty$. Now we must add  combinatoric factors: the diagram stays the same if we interchange the \emph{two} external legs attached to the vertex, so we add a $1/2$; and it is also invariant under a reflection through a vertical axis passing through the center of the loop, so we have another $1/2$:
\begin{equation}
	\int\frac{d^4k}{(2\pi)^4}\frac{1}{2}\frac{\frac{1}{2}\lambda}{k^2+i\epsilon} .
\end{equation}
Finally, because of the definition of the connected generating functional, we have a factor of $i$ and because of the form of the Taylor series expansion, we must add a $\phi^2$ for each vertex in the diagram, so the final expression for the first diagram is
\begin{equation}
	i\int\frac{d^4k}{(2\pi)^4}\frac{1}{2}\frac{\frac{1}{2}\lambda\phi^2}{k^2+i\epsilon} ,
\end{equation}
where the $\phi$ is defined as the functional derivative of the connected generating functional with respect to the external source. \newline
\indent Repeating the above reasoning, we arrive at the conclusion that the $n$-th diagram is represented by the expression
\begin{equation}
	i\int\frac{d^4k}{(2\pi)^4}\frac{1}{2n}\left(\frac{\frac{1}{2}\lambda\phi^2}{k^2+i\epsilon}\right)^{n} .
\end{equation}
Summing all of them, we obtain
\begin{eqnarray}
i\int \frac{d^4k}{(2\pi)^4}\sum_{n=1}^{\infty} \frac{1}{2n} \left( \frac{\frac{1}{2}\lambda\phi^2}{k^2 + i\epsilon} \right)^n .
\end{eqnarray}
Performing a Wick rotation, that is, making the change of variable $k^0\rightarrow ik^0$ of the time component of the momentum; we eliminate the pre-factor of $i$ and leave a minus sign. We also identify the infinite sum with the following Taylor series expansion of the logarithm:
\begin{equation}
	\ln(1-x)=-\sum_{n=1}^{\infty}\frac{x^n}{n} ,
\end{equation}
and dropping the $i\epsilon$,
\begin{eqnarray}
\nonumber & & i\int \frac{d^4k}{(2\pi)^4}\sum_{n=1}^{\infty} \frac{1}{2n} \left( \frac{\frac{1}{2}\lambda\phi^2}{k^2 + i\epsilon} \right)^n = \frac{1}{2}\int \frac{d^4 k}{(2\pi)^4} \ln \left( 1 + \frac{\lambda \phi^2}{2k^2} \right) =  \\
& & \frac{\Lambda^4}{64\pi^2} \ln \left(1 + \frac{\lambda\phi^2}{2\Lambda^2} \right) + \frac{\lambda\phi^2\Lambda^2}{128\pi^2} - \frac{\lambda^2\phi^4}{256\pi^2} \ln \left( 1 + \frac{2\Lambda^2}{\lambda\phi^2} \right).
\end{eqnarray}
In general, we set the second derivative of the potential, evaluated at $\phi=0$, equal to the (squared) renormalized mass $\mu$ of the particle:
\begin{equation}
	\frac{d^2V}{d\phi^2}\Big{|}_{\phi=0} = \mu^2 .
\end{equation}
and for the the particular case of a massless scalar  (which we will treat in the following section), we have
\begin{eqnarray}
\frac{d^2V}{d\phi^2}\Big{|}_{\phi=0} = 0 .
\label{2DV0}
\end{eqnarray}
Another renormalization condition that is employed concerns the fourth derivative of the potential which is equated to the coupling constant $\lambda$ 
\begin{eqnarray}
\frac{d^4V}{d\phi^4}\Big{|}_{\phi=M} = \lambda ,
\label{4DV}
\end{eqnarray}
where we evaluate $\phi$ at $M$ to avoid IR divergences. $M$ is called the \emph{renormalization mass} which is an arbitrary quantity.

\section{\label{sec:Massless}Massless Self Interacting Real Scalar Field}

The Lagrangian for a real massless self interacting scalar field model is of the form
\begin{eqnarray}
\mathcal{L} = \frac{1}{2}(\partial_\mu \phi)^2 + \frac{1}{2}A(\partial_\mu \phi)^2 - \frac{\lambda}{4!}\phi^4 - \frac{1}{2}B\phi^2 - \frac{1}{4!}C\phi^4 ,
\end{eqnarray}
where the terms containing $A$, $B$, and $C$ are the counterterms. Hence the one-loop-level effective potential is
\begin{eqnarray}
\nonumber V(\phi) &=& \frac{\lambda}{4!}\phi^4 + \frac{1}{2}B\phi^2+\frac{1}{4!}C\phi^4 \\ 
&+& i\int \frac{d^4k}{(2\pi)^4}\sum_{n=1}^{\infty} \frac{1}{2n} \left( \frac{\frac{1}{2}\lambda\phi^2}{k^2 + i\epsilon} \right)^n ,
\end{eqnarray}
the integral over the sum of the diagrams can be expressed as the logarithm
\begin{eqnarray}
&& \nonumber \frac{1}{2}\int \frac{d^4 k}{(2\pi)^4} \ln \left( 1 + \frac{\lambda \phi^2}{2k^2} \right) =  \frac{\Lambda^4}{64\pi^2} \ln \left(1 + \frac{\lambda\phi^2}{2\Lambda^2} \right) \\
&+& \frac{\lambda\phi^2\Lambda^2}{128\pi^2} - \frac{\lambda^2\phi^4}{256\pi^2} \ln \left( 1 + \frac{2\Lambda^2}{\lambda\phi^2} \right) .
\end{eqnarray}
The detailed derivation of the integral is found in appendix \ref{app:Integral1}. We can drop the first term because it vanishes in the limit of large $\Lambda$; also in this limit we have
\begin{equation}
	\ln \left( 1 + \frac{2\Lambda^2}{\lambda\phi^2} \right) = - \ln \left( \frac{\lambda\phi^2}{2\Lambda^2} \right) ,
\end{equation}
substituting the analytical solution of the logarithmic integral that we solve we now arrive at the following expression for the potential
\begin{equation}
	V(\phi) = \frac{\lambda}{4!}\phi^4 + \frac{1}{2}B\phi^2+\frac{1}{4!}C\phi^4 + \frac{\lambda\phi^2\Lambda^2}{128\pi^2} + \frac{\lambda^2\phi^4}{256\pi^2} \ln \left( \frac{\lambda\phi^2}{2\Lambda^2} \right),
\end{equation}
the counterterm coefficient $B$ is obtained through eq. \ref{2DV0}
\begin{equation}
	B=-\frac{\lambda\Lambda^2}{64\pi^2} ,
\end{equation}
and the counterterm coefficient $C$, from the condition given by eq. \ref{4DV}
\begin{equation}
	C=-\frac{\lambda^2\left[ 25+6\ln\left(\frac{\lambda M^2}{2\Lambda^2}\right)\right]}{64\pi^2},
\end{equation}
after substituting them into the effective potential, and a little bit of work, we get
\begin{equation}
	V(\phi)= \frac{\lambda}{4!}\phi^4+\frac{\lambda^2\phi^4}{256}\left[\ln\left(\frac{\phi^2}{M^2}\right)-\frac{25}{6}\right] .
\end{equation}
It is important to notice that although the coefficients of the counter terms $B$ and $C$ that we obtained are different to those obtained in Ref. \cite{Coleman:1973jx}, our final expression for the potential is the same.

\section{\label{sec:massive} Massive Self Interacting Real Scalar Field}

The Lagrangian for a massive self interacting scalar field theory is given by
\begin{eqnarray}
\nonumber \mathcal{L} &=& \frac{1}{2}(\partial_\mu \phi)^2 + \frac{1}{2}A(\partial_\mu \phi)^2 - \frac{\mu^2}{2}\phi^2 - \frac{\lambda}{4!}\phi^4 \\
&-& \frac{1}{2}B\phi^2 - \frac{1}{4!}C\phi^4 ,
\end{eqnarray}
so we identify the potential as
\begin{eqnarray}
\nonumber V(\phi) &=& \frac{\mu^2}{2}\phi^2 + \frac{\lambda}{4!}\phi^4 + \frac{1}{2}B\phi^2+\frac{1}{4!}C\phi^4 \\
&+& i\int \frac{d^4k}{(2\pi)^4}\sum_{n=1}^{\infty} \frac{1}{2n} \left( \frac{\frac{1}{2}\lambda\phi^2}{k^2 - \mu^2 + i\epsilon} \right)^n .
\end{eqnarray}
Once again we are left with an integral over the sum of the one loop level self interacting contributions which we write as a logarithmic integral (the detailed derivation is found in appendix \ref{app:Integral2})
\begin{eqnarray}
&& \nonumber \frac{1}{2}\int \frac{d^4k}{(2\pi )^4}\ln \left( 1+\frac{\lambda\phi^2}{2(k^2+\mu^2-i\epsilon)} \right)  = -2\mu^4\ln\left( \frac{\mu^2}{\Lambda^2} \right)\\
&+& \frac{1}{2}(2\mu^2+\lambda\phi^2)^2\ln\left(\frac{2\mu^2+\lambda\phi^2}{2\Lambda^2}\right)+\lambda\Lambda^2\phi^2 ,
\end{eqnarray}
from which we arrive at the effective one loop level potential for a massive interacting scalar field
\begin{eqnarray}
\nonumber && V(\phi)=\frac{\mu^2}{2}\phi^2+\frac{\lambda}{4!}\phi^4 \\
\nonumber &+& \frac{1}{64\pi^2}\left[ \left( \mu^2+\frac{\lambda\phi^2}{2}\right)^2 \ln \left( 1+\frac{\lambda\phi^2}{2\mu^2}\right) - \frac{1}{2}\lambda\mu^2\phi^2 \right. \\
	&-& \left. \frac{25}{24}\lambda^2\phi^4 + \frac{1}{4}\lambda^2\phi^4 \ln \left( \frac{2\mu^2}{\lambda M^2} \right) \right] .
\end{eqnarray}
The methodology used in obtaining the analytical solution for this case will now be extended to calculate the potential of interacting scalar fields in the THDM.

\section{\label{THDM}Effective Potential for a Two Higgs Doublet Model}

Because we are interested in the one-loop vacuum corrections of the scalar sector of the THDM (a review on the THMD can be found in Ref. \cite{BrancoReview}), we will only work with the contributions arising from the scalar potential
\begin{eqnarray}
\nonumber V(\Phi_1, \Phi_2) &=& m_1^2 \Phi_1^\dagger\Phi_1 + m_2^2 \Phi_2^\dagger\Phi_2 - m_{12}^2 \left(\Phi_1^\dagger \Phi_2 + \Phi_2^\dagger \Phi_1 \right) \\
\nonumber &+& \lambda_1 \left( \Phi_1^\dagger \Phi_1\right)^2 + \lambda_2 \left( \Phi_2^\dagger \Phi_2\right)^2 \\
\nonumber &+& \lambda_3 \left( \Phi_1^\dagger \Phi_1\right)\left( \Phi_2^\dagger \Phi_2\right) + \lambda_4 \left( \Phi_1^\dagger \Phi_2\right)\left( \Phi_2^\dagger \Phi_1\right) \\
&+& \lambda_5 \left[ \left( \Phi_1^\dagger \Phi_2\right)^2 + \left( \Phi_2^\dagger \Phi_1\right)^2 \right] ,
\end{eqnarray}
where the Higgs doublets can be written as
\begin{eqnarray}
{\Phi_1} =  \left( \begin{array}{c}
\phi_1^{+}  \\
\phi_1^0     
   \end{array} \right), \quad {\Phi_2} =  \left( \begin{array}{c}
\phi_2^{+}  \\
\phi_2^0
        \end{array} \right) ,
\end{eqnarray}
but the physical fields of the charged sector are defined through the transformation
\begin{eqnarray}
\left( \begin{array}{c}
G^{\pm}  \\
H^{\pm}     
   \end{array} \right) = \left( \begin{array}{c c}
\cos\beta & \sin\beta  \\
-\sin\beta & \cos\beta
        \end{array} \right) \left( \begin{array}{c}
\phi_1^\pm  \\
\phi_2^\pm
   \end{array} \right) ,
\end{eqnarray}
while the imaginary terms are transformed via
\begin{eqnarray}
\left( \begin{array}{c}
H^{0}  \\
h^{0}     
   \end{array} \right) = \left( \begin{array}{c c}
\cos\alpha & \sin\alpha  \\
-\sin\alpha & \cos\alpha
        \end{array} \right) \left( \begin{array}{c}
\mbox{Re} \phi_1^0  \\
\mbox{Re} \phi_2^0
   \end{array} \right) ,
\end{eqnarray}
and the real neutral sector is obtained with the transformation
\begin{eqnarray}
\left( \begin{array}{c}
G^{0}  \\
A^{0}     
   \end{array} \right) = \left( \begin{array}{c c}
\cos\beta & \sin\beta  \\
-\sin\beta & \cos\beta
        \end{array} \right) \left( \begin{array}{c}
\mbox{Im} \phi_1^0  \\
\mbox{Im} \phi_2^0
   \end{array} \right) .
\end{eqnarray}
Through the previously defined transformations we are able to obtain all of the self interaction vertices between each of the fields $h^0$, $H^0$, $A^0$, $H^\pm$, $G^0$ and $G^\pm$ found in Ref. \cite{Gunion:1989we}. But since we choose to work in the unitary gauge, the Goldstone bosons will be ``eaten'' by the SU(2)$\times$U(1) gauge fields. Therefore, we will only be dealing with the physical fields. 
\begin{figure}[H]
\center
\includegraphics[scale = 0.2]{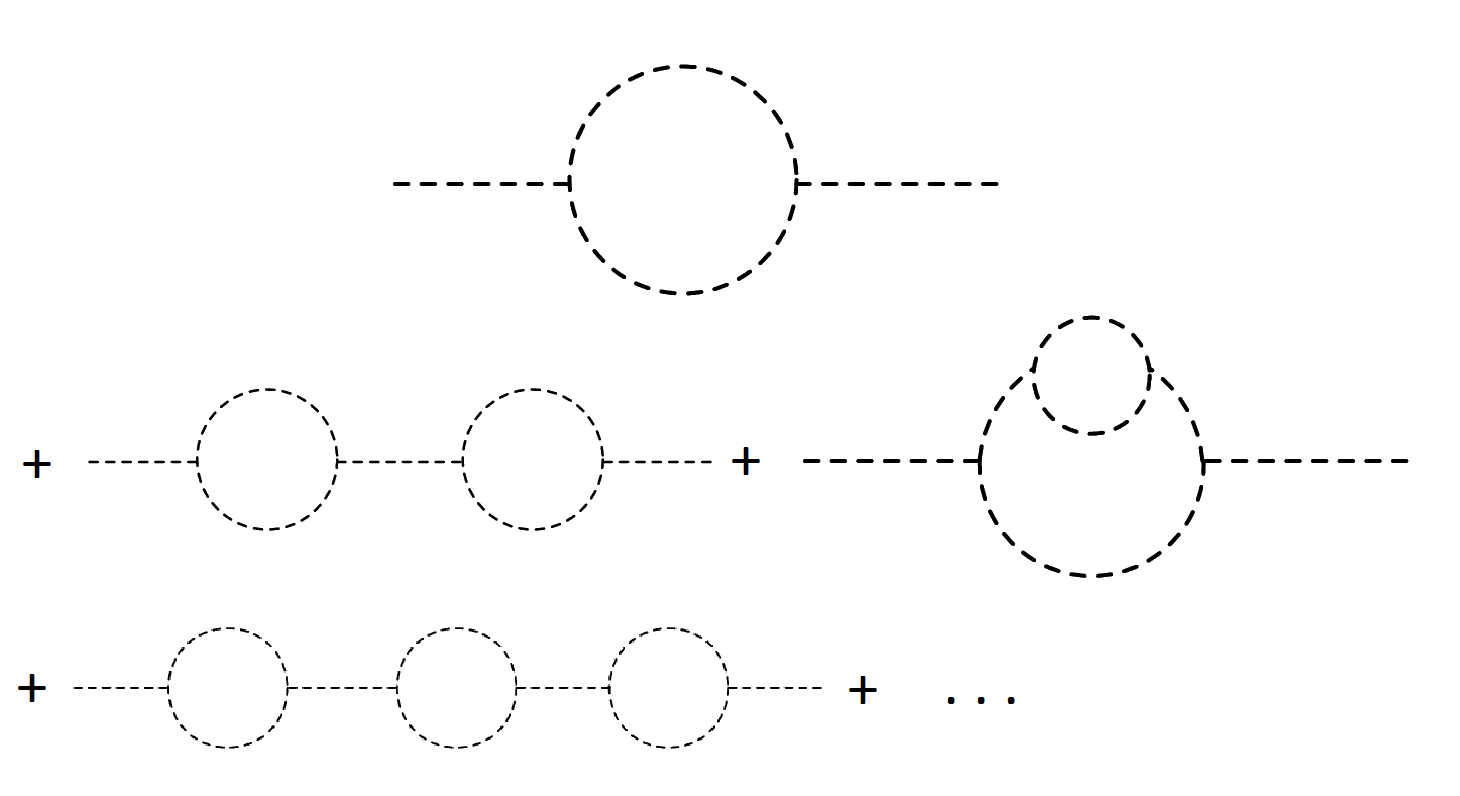}
\caption{Scalar propagator correction diagrams.
\label{fd:propagator}}
\end{figure}
Furthermore, we only consider quartic vertices. Diagrams containing triple vertices such as those shown in fig. \ref{fd:propagator} give loop corrections to the propagator, but they are not relevant to our present work, so we do not consider them. The effective potential in integral form is given in appendix \ref{app:Integral3}, from which we obtain the effective potential as a function of the physical fields:
 \begin{eqnarray}
\nonumber V(\Phi_1, \Phi_2) &=& V_{\mbox{tree}}\\
\nonumber &+& V_{H^{0}H^{0}H^{0}H^{0}} + V_{h^{0}h^{0}h^{0}h^{0}} + V_{A^{0}A^{0}A^{0}A^{0}} \\
\nonumber &+& V_{H^{0}h^{0}h^{0}h^{0}}   + V_{h^{0}H^{0}H^{0}H^{0}}+  V_{H^{0}H^{0}h^{0}h^{0}} \\
\nonumber                             &+& V_{H^{0}H^{0}A^{0}A^{0}}  + V_{h^{0}h^{0}A^{0}A^{0}} + V_{H^{+}H^{+}H^{-}H^{-}} \\
\nonumber &+&  V_{h^{0}h^{0}H^{-}H^{+}} + V_{H^{0}H^{0}H^{-}H^{+}} + V_{A^{0}A^{0}H^{-}H^{+}}  \\
                                              &+&  V_{A^{0}A^{0}H^{0}h^{0}} + V_{H^{0}h^{0}H^{-}H^{+}} ,
\end{eqnarray}
the first three terms corresponding to neutral Higgs of the same type $V_{SSSS}$ where $S=h0$, $H0$, $A0$ are given by
\begin{eqnarray}
\nonumber V_{SSSS} &=& - \lambda\mu_S^2 S^2  - \frac{25}{12} \lambda^2 S^4 + \frac{\lambda^2 S^4(16 \lambda \mu_S^2 M^2 - 64 \mu_S^4)}{12 (\lambda M^2 + 2 \mu_S^2)^2} \\
\nonumber &+& 2\lambda \mu_S^2 S^2 \ln \left(1 + \frac{\lambda S^2}{2\mu_S^2} \right) + 2\mu_S^4\ln \left(1 + \frac{\lambda S^2}{2\mu_S^2} \right) \\
&+& \frac{1}{2}\lambda^2 S^4 \ln \left(\frac{\mu_S^2 + \frac{\lambda S^2}{2}}{\mu_S^2 + \frac{\lambda M^2}{2}} \right) ,
\end{eqnarray}        
next we have the coupling of three fields of the same type ($S_1$) with another one of a different kind ($S_2$) given by        
         \begin{eqnarray}
\nonumber V_{{S_1}{S_1}{S_1}{S_2}} &=& - \lambda\mu_{S_1}^2 S_2 S_1 \left( 1 - 2 \ln\left(1 + \frac{\lambda S_2 S_1 }{2 \mu_{S_1}^2} \right) \right) \\
\nonumber &+& \frac{\lambda^2}{4} S_2^2 S_1^2 \left( 2\ln\left( 1 + \frac{\lambda S_2 S_1}{2 \mu_{S_1}^2} \right) - 3\right) \\
&+& 2\mu_{S_1}^4\ln \left( 1 + \frac{\lambda S_2 S_1}{2\mu_{S_1}^2}\right) ,
        \end{eqnarray}
during the calculation for this case, we set the one of the mass counter term coefficients $B=0$, because if one includes all the mass counter terms you end up with a divergent logarithm. 

The subsequent four terms of the effective potential which have two equal pairs $S_1$, $S_2$ are given by
        \begin{eqnarray}
\nonumber  V_{S_1S_1S_2S_2} &=& - \lambda \mu_S^2(S_1^2 + S_2^2) - \frac{3}{4}\lambda^2 (S_1^4 + S_2^4) \\
        &+& \frac{4(S_1^4 + S_2^4)M^2\lambda^3}{3(M^2\lambda + 2\mu_S^2)^2}(\mu_S^2 - M^2\lambda) \nonumber \\
        &+& 2\lambda \mu_S^2 S_1^2\ln \left( 1 + \frac{\lambda S_1^2}{2\mu_S^2}\right) + 2\lambda \mu_S^2 S_2^2\ln \left( 1 + \frac{\lambda S_2^2}{2\mu_S^2}\right) \nonumber \\
\nonumber &+& \frac{\lambda^2}{2}S_1^4 \ln \left( \frac{\mu_S^2 + \frac{\lambda S_1^2}{2}}{\mu_S^2 + \frac{\lambda M^2}{2}}\right) \\
&+& \frac{\lambda^2}{2}S_2^4 \ln \left( \frac{\mu_S^2 + \frac{\lambda S_2^2}{2}}{\mu_S^2 + \frac{\lambda M^2}{2}}\right).
                \end{eqnarray}
Finally we have the terms in which we have a single pair of scalars of the same kind and the remaining two are of a different type
        \begin{eqnarray}
\nonumber        V_{S_1S_1S_2S_3} &=& - \lambda\mu_{S_1}^2 S_2 S_3 \left( 1 - 2 \ln\left(1 + \frac{\lambda S_2 S_3 }{2 \mu_{S_1}^2} \right) \right) \\
\nonumber &+& \frac{\lambda^2}{4} S_2^2 S_3^2 \left( 2\ln\left( 1 + \frac{\lambda S_2 S_3}{2 \mu_{S_1}^2} \right) - 3\right) \\
&+& 2\mu_{S_1}^4\ln \left( 1 + \frac{\lambda S_2 S_3}{2\mu_{S_1}^2}\right) .
        \end{eqnarray}
the last term in the potential involves the coupling of fields which are all of a different kind, such as $H^0h^0H^+H^-$, this type of coupling does not permit a polygon diagram whose series is constructible.
\section{\label{conclusion} Conclusions}\label{sec:conclusion}

We calculated the contributions of the THDM effective potential that come from one-loop diagrams with quartic coupling vertices which are dimensionally consistent and renormalizable. We hope this analytical scheme will be of use to model builders interested in theories that must take in to consideration vacuum corrections to two or more scalars.
\begin{acknowledgments}
We acknowledge support from CONACYT (M\'exico), and appreciate the advice and discussions from Gerardo F. Torres del Castillo and O. Meza-Aldama.
\end{acknowledgments}


\section{Appendix}

\subsection{Solution to the integral of the massless effective potential} \label{app:Integral1}

We proceed to solve the integral by first changing from an integral over 4-momentum to an integral over hyper spherical coordinates

\begin{eqnarray}
\nonumber k^0   &=& \rho \cos\psi, \\ \nonumber
k^1   &=& \rho \sin\psi \cos\theta, \\ \nonumber
k^2   &=& \rho \sin \psi \sin\theta\cos\phi, \\ \nonumber
k^3   &=& \rho \sin\psi\sin\theta\sin\phi, \\ \nonumber
d^4k &=& \rho^3\sin^2 \psi \sin\theta d\rho d\psi d\theta d\phi ,
\end{eqnarray}

where $\theta$, $\psi$ $\epsilon$ $(0,\pi)$ and $\phi$ $\epsilon$ $(0,2\pi)$.

From which we obtain

\begin{eqnarray}
\nonumber && \frac{1}{2}\int \frac{d^4k}{(2\pi)^4}\ln\left(1+\frac{\lambda\phi^2}{2k^2} \right)= \frac{1}{2(2\pi)^4}\int_0^\pi \sin^2\psi d\psi \\
&& \int_0^\pi \sin\theta d\theta\int_0^{2\pi} d\phi \int_0^\Lambda \rho^3 \ln \left(1 + \frac{\lambda \phi^2}{2\rho^2} \right) d\rho . 
\end{eqnarray}

We now solve the integral by

\begin{eqnarray}
\nonumber I &=& \frac{1}{16\pi^4} \int_0^\Lambda \rho^3 \ln\left( 1 + \frac{\lambda\phi^2}{2\rho^2}\right) d\rho \\
\nonumber &=& \frac{1}{16 \pi^2} \int_0^\infty \ln \left( 1 + \frac{\lambda\phi^2}{2\rho^2}\right) d\left( \frac{1}{4} \rho^4\right) \\ \nonumber
 &=& \frac{1}{16\pi^2} \left[\frac{1}{4} \rho^4 \ln\left(1 + \frac{\lambda\phi^2}{2\rho^2} \right) \Big{|}_0^\Lambda -  \frac{1}{4} \int_0^\Lambda \rho^4 \frac{1}{1+\frac{\lambda\phi^2}{2\rho^2}} (-\frac{\lambda\phi^2} {\rho^3})\right]\\
 &=&\frac{\Lambda^4}{64\pi^2} \ln \left( 1 + \frac{\lambda\phi^2}{2\Lambda^2}\right) + \frac{1}{64\pi^2} \int_0^\Lambda \rho^4\frac{2\rho^2}{2\rho^2 + \lambda\phi^2} \frac{\lambda\phi^2}{\rho^3} d\rho ,
\end{eqnarray}
the first term cancels because when $\Lambda\to\infty$ the logarithm goes to zero, as for the last term 
\begin{eqnarray}
I_1= \frac{\lambda\phi^2}{32\pi^2} \int_0^\Lambda \frac{\rho^3 d\rho}{2\rho^2+\lambda\phi^2} ,
\end{eqnarray}

making $x=2\rho^2\Longrightarrow$ $dx=4\rho$ $d\rho\Longrightarrow$ $xdx=8\rho^3d\rho$,

\begin{eqnarray}
I_1=\frac{\lambda\phi^2}{256\pi^2} \int_0^{2\Lambda^2} \frac{xdx}{x+\lambda\phi^2} .
\end{eqnarray}

Now defining $y=x + \lambda\phi^2$ $\Longrightarrow$ $dy=dx$

\begin{eqnarray}
I_1 &=& \frac{\lambda\phi^2}{256\pi^2} \int_{\lambda\phi}^{2\Lambda^2 + \lambda\phi^2} \frac{y-\lambda\phi^2}{y}dy \\
\nonumber &=& \frac{\lambda\phi^2}{256\pi^2}\int_{\lambda\phi}^{2\Lambda^2 + \lambda\phi^2}  \left( 1 - \frac{\lambda\phi^2}{y}\right) dy \\
\nonumber &=& \frac{\lambda\phi^2}{256\pi^2} \left[ 2\Lambda^2 - \lambda\phi^2 \ln\left( \frac{2\Lambda^2 + \lambda\phi^2}{\lambda\phi^2} \right) \right] \\
&=& \frac{\lambda\phi^2\Lambda^2}{128\pi^2} - \frac{\lambda^2\phi^4}{256\pi^2}\ln\left( 1+ \frac{2\Lambda^2}{\lambda\phi^2}\right) .
\end{eqnarray}
In the large cutoff limit, $\Lambda \gg 1$, we have
\begin{equation}
	\ln\left( 1+\frac{2\Lambda^2}{\lambda\phi^2}\right)=-\ln\left(\frac{\lambda\phi^2}{2\Lambda^2}\right) ,
\end{equation}
so the effective potential is
\begin{equation}
	V = \frac{\lambda}{4!}\phi^4 + \frac{1}{2}B\phi^2+\frac{1}{4!}C\phi^4+\frac{\lambda\phi^2\Lambda^2}{128\pi^2} + \frac{\lambda^2\phi^4}{256\pi^2}\ln\frac{\lambda\phi^2}{2\Lambda^2} .
\end{equation}
\subsection{Solution to the integral of the massive effective potential}  \label{app:Integral2}

In this case we have that the sum of 1-loop diagrams is given by the following integral:
\begin{eqnarray}
\nonumber I &\equiv& \frac{1}{2} \int \frac{d^4 k} {(2\pi)^4} \ln( 1 + \frac{\lambda \phi^2}{2(k^2 + \mu^2 - i\epsilon)}) \\
&=& \frac{1}{2}(2 \pi^2) \frac{1}{(2\pi)^4} \int \rho^3 \ln( 1 + \frac{\lambda \phi^2}{2(\rho^2 + \mu^2)})d \rho ,
\end{eqnarray}
we now make the change of variable $x = 2 \rho^2 \Rightarrow$ $dx=4\rho d \rho$ $\Rightarrow$ $xdx=8\rho^3 d \rho$. Then dropping the $i\epsilon$,
\begin{eqnarray}
\nonumber I & = & \frac{1}{16\pi^2} \frac{1}{8} \int x \ln (1 + \frac{\lambda\phi^2}{x + \mu^2})d x \\
\nonumber &=& \frac{1}{16\pi^2} \frac{1}{8} \int \ln (1 + \frac{\lambda\phi^2}{x + \mu^2})d\left( \frac{1}{2} x^2 \right) \\
\nonumber &=& \frac{1}{128\pi^2} \frac{1}{2}x^2 \ln (1 + \frac{\lambda\phi^2}{x + 2\mu^2}) \\
\nonumber && - \frac{1}{128\pi^2} \int \frac{1}{2} x^2 \frac{1}{1 + \frac{\lambda \phi^2}{x + 2\mu^2}}\frac{-\lambda \phi^2}{(x + 2\mu^2)}dx \\
\nonumber &=& \frac{1}{256\pi^2} x^2 \ln (1 + \frac{\lambda \phi^2}{x + 2\mu^2}) \\
&& + \frac{\lambda \phi^2}{256\pi^2} \int \frac{x^2 dx }{(x + 2\mu^2 + \lambda \phi^2)(x + 2\mu^2)} .
\end{eqnarray}
Let's now solve the rightmost integral (redefining some $x$-independent quantities as $a$ and $b$):
\begin{equation}
I_1 = \int \frac{x^2 dx}{(x + a)(x + b)} ,
\end{equation}
setting $u=x^2$ and $dv=\frac{dx}{(x + a)(x + b)}$, to get $v$ we integrate:
\begin{eqnarray}
\nonumber v &=& \int \frac{dx}{(x + a)(x + b)} = \int \frac{dy}{y(y - a + b)} \\
&=& \frac{-1}{-a + b} \ln (\frac{b - a + y}{y}) .
\end{eqnarray}
We now use the following known integral
\begin{equation}
\int \frac{du}{u(A + Bu)} = -\frac{1}{A} \ln(\frac{A + Bu}{u}) ,
\end{equation}
where in our case $B =1$, $A=-a + b$, and $u=y$; thus
\begin{eqnarray}
\nonumber I_1 &=& \int \frac{x^2 dx}{(x + a)(x + b)} \\
\nonumber &=& \frac{x^2}{a-b} \ln (\frac{b + x}{a +x}) - \int \frac{1}{a - b} \ln (\frac{b + x}{a + x})2x dx \\
\nonumber &=& \frac{x^2}{a - b} \ln (\frac{x + b}{x +a})  \\
&& + \frac{2}{b - a} \int \left[x\ln(x+b) - x\ln(x + a) \right] dx .
\end{eqnarray}
For each term on the right we have
\begin{equation}
	\int x \ln (x + b) dx = \frac{1}{2} (x^2 - b^2) \ln (x+b) - \frac{1}{4}(x - b)^2 + b^2 ,
\end{equation}
\begin{equation}
\int x \ln (x + a) dx = \frac{1}{2}(x^2 - a^2) \ln (x +a) - \frac{1}{4} (x-a)^2 + a^2 ,
\end{equation}
putting both together, and doing some algebra,
\begin{equation}
\nonumber I_1 = \frac{b^2}{a-b} \ln(x+b) + \frac{a^2}{b-a} \ln (x +a) + x + \frac{3}{2} (b + a) .
\end{equation}
Finally, placing a cutoff at $k^2=\Lambda^2$, or equivalently $x^2=2\Lambda^2$,
\begin{eqnarray}
\nonumber I &=& \frac{1}{2} \int \frac{d^4 k}{(2\pi)^4} \ln \left(1 + \frac{\lambda \phi^2}{2(k^2 + \mu^2)} \right) \\
\nonumber &=& \frac{1}{256\pi^2} x^2 \ln \left(1 + \frac{\lambda \phi^2}{x + 2\mu^2} \right) \Big|^{2\Lambda^2}_0 \\
\nonumber && + \frac{\lambda \phi^2}{256\pi^2} \left[ \frac{1}{\lambda \phi^2}\left[ 4\mu^4\ln(x+2\mu^2) \right.\right. \\
\nonumber && \left. - (2\mu^2 + \lambda\phi^2)^2\ln(x + 2\mu^2 + \lambda\phi^2)\right] + x \\
&& + \left. \frac{3}{2}(\lambda\phi^2 + 4\mu^2)\right]\Big|^{2\Lambda^2}_0 .
\end{eqnarray}
We neglect the first term since when $\Lambda \to \infty$ it tends to zero. Then, evaluating, we're left only with
\begin{eqnarray}
\nonumber	I &=& \frac{1}{256\pi^2} \left[ 4\mu^4\ln\left( 1+\frac{\Lambda^2}{\mu^2}\right)\right. \\
\nonumber	&& - \left. (2\mu^2+\lambda\phi^2)^2\ln\left( 1+\frac{2\Lambda^2}{2\mu^2+\lambda\phi^2}\right) \right] \\
	&& + \frac{\lambda\phi^2\Lambda^2}{128\pi^2} .
\end{eqnarray}
In the limit of large $\Lambda$, we get
\begin{eqnarray}
\nonumber	I &=& \frac{1}{128\pi^2}\left[ -2\mu^2\ln\left( \frac{\mu^2}{\Lambda^2}\right) \right. \\
	&+& \left. \frac{1}{2} (2\mu^2+\lambda\phi^2)^2 \ln\left( \frac{2\mu^2+\lambda\phi^2}{2\Lambda^2}\right) + \lambda\Lambda^2\phi^2 \right] .
\end{eqnarray}

\subsection{THDM intergral form effective Potential}  \label{app:Integral3}

There are essentially three types of vertices involved in the calculation: those with four equal fields, with two pair of equal fields, and with only one pair of equal fields. We only consider the case in which only one particle goes around the loop. Thus, when we analyze a vertex with more than one particle interacting, we write only the propagators of the particle that is ``repeated''. In the first case, the integral form of the potential is
\begin{equation}
	V_{SSSS}=V_{\mbox{tree}}+\int \frac{d^4k}{(2\pi)^4} \ln \left( 1+ \frac{\lambda S^2}{2(k^2+\mu_S^2)} \right) ,
\end{equation}
where $\lambda$ represents genericallly the self-coupling constant and $\mu_S$ is the mass of the particle $S$, and we've omitted the $i\epsilon$. In the second case,
\begin{equation}
	V_{S_1S_1S_2S_2}=V_{\mbox{tree}}+\int \frac{d^4k}{(2\pi)^4} \ln \left( 1+ \frac{\lambda S_1^2}{2(k^2+\mu_{S_2}^2)} \right) ,
\end{equation}
where there's an obvious symmetry between $S_1$ and $S_2$. Finally, for the last case,
\begin{equation}
	V_{S_1S_1S_2S_3}=V_{\mbox{tree}}+\int \frac{d^4k}{(2\pi)^4} \ln \left( 1+ \frac{\lambda S_2S_3}{2(k^2+\mu_{S_1}^2)} \right) .
\end{equation}





\end{document}